\documentclass[aps,prl,amsmath,twocolumn]{revtex4}
\usepackage{bm}
\usepackage{graphicx}
\usepackage{amssymb}
\usepackage{subcaption}
\usepackage{multirow}
\usepackage{makecell}
\def \d{\partial}

\def \be{\mathbf{e}}

\begin{document}

\title{Stochastic identities for random isotropic fields}

\author{ A.S. Il'yn$^{1,2}$, A.V. Kopyev$^1$, V.A. Sirota$^1$, K.P. Zybin$^{1,2}$
\thanks{Electronic addresses:  asil72@mail.ru, kopyev@lpi.ru, sirota@lpi.ru, zybin@lpi.ru}}
\affiliation{$^1$ P.N.Lebedev Physical Institute of RAS, 119991, Leninskij pr.53, Moscow,
Russia \\
$^2$ National Research University Higher School of Economics, 101000, Myasnitskaya 20, Moscow,
Russia}

\begin{abstract}
This letter presents new nontrivial stochastic identities for random isotropic second rank tensor
fields. They can be considered as markers of statistical isotropy in turbulent flows of any
nature. The case of axial symmetry is also considered.   We confirm the validity of the identities
using different direct numerical simulations of turbulent flows.
%\\
%\vspace{0.4cm} \noindent {\small PACS numbers: }
\end{abstract}

\maketitle

\section{ Introduction}

It is known that the flow of viscous fluid is unstable at high Reynolds numbers. In
three-dimensional hydrodynamic flow after some time
%at some point after the beginning of its evolution,
the velocity field becomes chaotic and the spectrum of its pulsations extends from the integral to
the dissipative (Kolmogorov) scale \cite{K41, Frisch}. This phenomenon is called fully developed
turbulence. One of Kolmogorov's hypotheses states that random differences in velocity fluctuations
at two arbitrary points $\delta \mathbf{u}$ in such a flow are statistically homogeneous and
%, moreover,
isotropic if the points are located quite far from the boundaries of the flow~\cite{MY75}. The
verification of the hypothesis in a specific stream remains a challenge~\cite{PRFl-iso}.

From mathematical point of view, the isotropy of the flow means that the probability measure of
$\delta \mathbf{u}$ is invariant with respect to the group of space motions  (the semi-direct
product of parallel transport  and orthogonal transformations $\mathbb{R}\rtimes O(3)$):
$$
\mathbf{u}\bigl(t,\mathbf{r}\bigr)\rightarrow \mathbf{u}\bigl(t, \mathbf{R}^{-1}(\mathbf{r}-\mathbf{a})
+\mathbf{a}\bigr),\,\,\,\,\mathbf{a}\in\mathbb{R}^3 \ , \mathbf{R}\in O(3)
$$
Let us now consider tensor fields that are functionally dependent on the velocity field. For
instance, it can be a velocity gradients field usually studied in the theory of
turbulence~\cite{Meneveau, JohnWilc}:
$$A_{kp}=\frac{\partial u_k}{\partial r_p} \ , $$
or a magnetic induction gradients field in magneto-hydrodynamic turbulence~\cite{Schek, KuzMikh}:
$$H_{kp}=\frac{\partial B_k}{\partial r_p} \ ,$$
or a field of second derivatives of a scalar ~$\phi$, which is studied in the theory of advection
of scalar fields~\cite{FGV, Dimotekis, Foux}:
$$\Phi_{kp}=\frac{\partial^2 \phi}{\partial r_k \partial r_p}$$

In the absence of external factors that could affect the field dynamics, these advected fields
become statistically (locally) isotropic in a fully developed turbulent flow; all the fields have
finite correlation lengths, and unlike, e.g., ferromagnetics, there can be no spontaneous symmetry
breaking in turbulent media.

So, local isotropy is an inherent property of fully developed turbulence, even though statistical
homogeneity and stationarity can be violated.  If the flow is, in addition, homogenous and/or
stationary, then the corresponding ergodic theorem is valid, and
%
%the probability measure of such tensors is the same at each point and invariant with respect to
%orthogonal transformations. A natural assumption is that there is no spontaneous breaking of
%rotational symmetry in a turbulent flow (unlike in, for example, ferromagnets), so all the fields
%have a finite correlation length. This means that
the mathematical expectations of every function of the field can be calculated or measured by
taking spatial or time average.

Isotropy leads to the existence of the so-called stochastic identities, i.e. some functions of the
tensor components with universal mathematical expectations that do not change during evolution. The
simplest example of such an identity is the following:
\begin{equation} \label{0}
\left \langle \left( \frac{\delta u_{1}}{\delta u} \right)^2 \right \rangle= \left \langle \left(
\frac{\delta u_{2}}{\delta u} \right)^2 \right \rangle= \left \langle \left( \frac{\delta
u_{3}}{\delta u} \right)^2 \right \rangle=\frac13
\end{equation}
This identity is
trivial. Moreover, it remains true not only for isotropic fields, but also for the fields with
probability measure invariant only under coordinate axes permutations (and not isotropic). 

This letter presents far less trivial stochastic identities that can be used as markers of
isotropy for arbitrary second rank tensors. We focus on the  three dimensional case; however, they
can as well be applied to spaces of any dimension.

 \section{ LDU decomposition of matrices}

The human experience shows that when trying to better understanding of a statement it should be formulated in a broadest way possible. Therefore, it would not hurt to consider a Euclidean
space with arbitrary dimension~$d$.

Consider arbitrary second rank tensors $\mathbf{A}$ and symmetric positive-definite quadratic form $\mathbf{\Gamma}=\mathbf{A}^T \mathbf{A}$. In accordance with the
Gauss (lower-diagonal-upper, LDU) decomposition, in the given basis its matrix can be presented in
the form:
%
%We set $\mathbf{\Gamma}$ to be a real, symmetric, positive definite $d\times d$-matrix. Consider its Jacobi decomposition,
\begin{equation}\label{E:Gamma}
\mathbf{\Gamma}=\mathbf{Z}^T \mathbf{D}^2 \mathbf{Z},
\end{equation}
where $\mathbf{D}=diag\{D_1,\dots,D_d \}$, $D_i > 0$, $\mathbf{Z}$ is an upper-triangular matrix
with units on the main diagonal.
%This presents the diagonalization of  $\mathbf{\Gamma}$ by means of a triangular matrix.
%The set $\{ D_i \}$ is  called Jacobi coefficients

It is important to note that the coefficients $D_i$ are not equal to  the eigenvalues of
$\mathbf{\Gamma}$, except for the case of a diagonal matrix. %
Unlike the eigenvalues, they are not universal characteristics of the quadratic form: the set $\{
D_i \}$ depends on the chosen basis.

 It follows from~(\ref{E:Gamma}) that
\begin{equation}\label{E:Jacobi}
D_k^2=s_k^2/s_{k-1}^2,
\end{equation}
where $s_k^2$ is the $k$-order leading principal minor of $\boldsymbol{\Gamma}$.

\section{ Symmetry properties of $\{ D_i \}$ for isotropic random tensors}

Let now $\mathbf{\mathbf{A}}$ be random, and
%
%Consider a random matrix $\mathbf{\Gamma}$,
let its probability measure be invariant under the
rotation in the $p,p+1$ plane where $p\in\{1,\dots,d-1\}$. This means that the probability density
$\rho(\mathbf{\mathbf{A}})$ satisfies the identity
\begin{equation}\label{E:p-iso}
\rho(\mathbf{\mathbf{A}})=\rho(\mathbf{O}_p^{-1} \mathbf{\mathbf{A}}\mathbf{O}_p ),
\end{equation}
where $\mathbf{O}_p$ is a rotation in the $p,p+1$ plane. Let $m_1,\cdots,m_d$ be an arbitrary set
of real numbers. Then, as it was shown in \cite{SIKZ}, the correlator
\begin{equation}\label{E:correlator}
C(m_1,\dots,m_d )=\langle D_1^{m_1+1}\cdots D_d^{m_d+d}\rangle
\end{equation}
 is symmetric with respect to the pair permutation $m_p\leftrightarrow m_{p+1}$.
This is a consequence of some nontrivial symmetric property of elliptic integrals, which appear as a result of integration over the SO(2) subgroup of SO(d).

  Any permutation $\{\pi(k)\}$ can be represented as a composition of pair permutations of adjacent
  numbers; % it follows that
  thus, for isotropic probability density
\begin{equation}
\rho(\mathbf{\mathbf{A}})=\rho(\mathbf{O}^{-1} \mathbf{\mathbf{A} O}), \forall \mathbf{O}\in O(d) \ ,
\end{equation}
the correlator~(\ref{E:correlator}) is symmetric with respect to any permutation
$m_1,\dots,m_d \rightarrow m_{\pi(1)},\dots,m_{\pi(d)}$.

\section{ Stochastic identities }

%As previously shown,
%C(m_1,…,m_d )=C(〖m'〗_1,…,〖m'〗_d ), for 〖m'〗_k=m_(π(k)),
%where π(k) is an arbitrary permutation of delements.
Substituting $m_k=-k$ in~(\ref{E:correlator}), one obtains
\begin{equation}  \label{7}
1=\langle D_1^{1-\pi(1)} \dots D_d^{d-\pi(d)} \rangle \ \   \forall \{ \pi(k) \}
\end{equation}
%From (\ref{E:Jacobi}), it follows similar identities for $s_k$:
By means of~(\ref{E:Jacobi})  we express~(\ref{7})  in terms of  the leading principal minors:
\begin{equation}\label{E:idents}
\langle s_1 ^{\pi(2)-\pi(1)-1}\dots s_{d-1}^{\pi(d)-\pi(d-1)-1} s_d^{d-\pi(d)}\rangle=1, \forall \pi(k)
\end{equation}
These identities are valid for arbitrary isotropic probability measures at any given time. Their
existence is entirely % only exist
due to the geometric properties of $O(d)$, regardless of the dynamics of the field.

From geometric point of view, $s_k$ describe the transformation of infinitesimal
segment/square/hypersquare... under the action of  $\bf A$. So, the identities can be interpreted
as (stochastic) relations between transformations of hypervolumes of different dimensions (see
\cite{SIKZ} and Supplemental materials for more detailed consideration).
%  Also, in \cite{IKSZ} a physical interpretation of $s_k$ in terms of $k$-dimensional surface densities was proposed.

Now, we focus on the most practically important case $d=3$. From~(\ref{E:idents}) we get
 five nontrivial identities (as many as there are nontrivial permutations):

\begin{table}[ht]
\caption{Stochastic identities for isotropic three-dimensional flows. The first two identities are
valid for flows with stochastic axial symmetry (1st and 3rd axis, respectively). }
  \centering
  \begin{tabular}{|l|l|}
        \hline
        \textbf{Permutation} & \textbf{Stochastic identity} \\
        \hline
        132 & $\langle s_1 s_2^{-2} s_3\rangle=1$   \\
        \hline
        213 & $\langle s_1^{-2} s_2\rangle=1$  \\
        \hline
        321 & $\langle s_1^{-2} s_2^{-2} s_3^2\rangle=1$  \\
        \hline
        231 & $\langle s_2^{-3} s_3^2\rangle=1$  \\
        \hline
        312 & $\langle s_1^{-3} s_3 \rangle=1$ \\
        \hline
  \end{tabular}
  %\caption{Table 1: In-text Table}
  \label{tab:1}
\end{table}

We note that symmetry of~(\ref{E:correlator}) and the relations~(\ref{7}),~(\ref{E:idents}) are not
just a result of the change of the axes numbers (as in~(\ref{0})). Actually, such a change of
numbers would not just exchange the values of $D_i$: it would change them essentially. The
existence of the identities is due to the existence of continuous (not discrete) symmetry.

On the other hand, since the set $\{ D_i \}$ is not invariant under the rotation of the basis, the %
identities~(\ref{E:idents})
%left side in (\ref{E:idents}) depends on the choice of the coordinate frame
%the left-side expressions of (\ref{E:idents})
written in different coordinate frames are essentially different: %
changing the frame, one gets new identities. %
Namely,  let us choose some basis and write down the identities~(\ref{E:idents}).  Let us now take
some rotated basis and write one of the identities~(\ref{E:idents}) in this new coordinate frame.
Then we express the components ${\Gamma}'_{ij}$ of the 'new' matrix in terms of the first matrix
$\mathbf{\Gamma}$, and the obtained identity is functionally independent on the previous five
 (in the sense that the expressions under the averaging brackets are linearly
 independent, and
there is no  relation between them that would be valid for all realizations).
% do not coincide  for almost any particular realization).
This way we get five continuous three-parametric (by the number of rotations) sets of stochastic
identities describing the isotropic random tensor.
% The number of parameters corresponds to the number of rotations of the three-dimensional space.

Returning to the case of arbitrary dimension,  we find $(d!-1)$  continuous families of identities,
each of them being $d(d-1)/2$ -parametric. Geometrical meaning of these families is discussed in Supplemental Material~1

Below, we illustrate this idea by  applying it to
%One more interesting  illustration to  this idea appears in
the case of axially symmetric flow.

\section{ Axial symmetry  }

Axisymmetric flows are found in  turbulent jets~\cite{kopiev}, channel flows~\cite{nikitin}, jets that are observed in
astrophysical objects~\cite{beskin}, etc.

 Consider a random tensor  that is not completely isotropic but has axial stochastic symmetry, i.e.,
 its
probability measure is invariant with respect to rotation around one direction. Choose the
coordinate frame in such a way that the selected direction coincides with the first axis. As
follows from the reasoning between~(\ref{E:p-iso})~and~(\ref{E:correlator}), the first identity in Table~1
still holds in this case ($p=2$ in~(\ref{E:p-iso})).  Thus, we get
$$
\left \langle s_1 s_2^{-2} s_3 \right \rangle = 1
$$
Now we take, e.g., the  coordinate frame with the basis vectors ${\bf e}'_1={\bf e}_1$, ${\bf e}'_2
=  {\bf e}_3$, ${\bf e}'_3 = -{\bf e}_2$. We can express the new components  ${\Gamma}'_{ij}$ in
terms of the primary frame: then, $s'_1=s_1$, $s'_3 = s_3$, $s_2'^2 = m_{13}$ where $m_{ij}$ is the
principal 2-rank minor obtained from $\mathbf{\Gamma}$ by preserving the rows and columns with
numbers $i,j$. Axis 1' is still the axis of symmetry, so we obtain one more equality:
$$
\left \langle s'_1 s_2'^{-2} s'_3 \right \rangle = \left \langle s_1  s_3 / m_{13} \right \rangle =
1 ,
$$
which contains principal and not only leading principle minors.

Rotating the frame by different angles around the first axis, one would get new identities; they
would contain different combinations of the components of $\Gamma$ (not only minors).

Furthermore, we take one more coordinate frame in such a way that the axis of symmetry is the third
coordinate axis: say, ${\bf e}''_1={\bf e}_2$, ${\bf e}''_2={\bf e}_3$, ${\bf e}''_3= {\bf e}_1$.
Then
 the same consideration of~(\ref{E:p-iso}) with $p=1$ shows that the second identity of Table~1 is
 valid. Returning back to the primary frame we get
$$
\left \langle s_{1}''^{-2} s''_2 \right \rangle = \left \langle (\Gamma_{22})^{-1} \sqrt{m_{23}}
\right \rangle = 1 ,
$$
The rotation of the second coordinate frame around the direction of symmetry (i.e., its third axis)
produces one more continuous family of identities.
%~\footnote{Choosing the second axis to be the distinguished direction, we
% would not get new identities,
%because the axial symmetry with respect to the second axis is not that of %the type
%of~(\ref{E:p-iso}),
%%no symmetry of the type~(\ref{E:p-iso}) corresponds to the rotation around
%% the second axis,
%and thus, no one of the identities listed in Table~1 holds in this case.}

Now, let ${\bf e}_2$ be the direction of axial symmetry. No one of the identities listed in Table~1
holds in this case, since both rotations ${\bf O}_1$ and ${\bf O}_2$ violate the symmetry. However,
choosing the coordinate frame $\{ \bf e \}''_i$ we get the axial symmetry about the new $x''$ axis,
so
$$
\left \langle s_{1}'' {s''_2}^{-2} s''_3 \right \rangle = \left \langle \sqrt{\Gamma_{22}}
m_{23}^{-1} s_3 \right \rangle = 1
$$
In the frame with basis vectors $\{ \bf e \}'_i$,  the direction of symmetry becomes the $z'$ axis, and the second line in Table~1 comes into play:
$$
\left \langle {s'_{1}}^{-2} {s'_2} \right \rangle = \left \langle s_1^{-2}  \sqrt{m_{13}} \right
\rangle = 1
$$
Of course, for each of these cases, one can additionally rotate the coordinate frame around the symmetry axis, and obtain more identities.

So, for any axially symmetric flow there are two continuous one-parametric families of identities.

We would like to note the difference between rotation of the coordinate frame and rotation
$O_p$ in (\ref{E:p-iso})-(\ref{E:correlator}). In whatever frame the invariance  (\ref{E:p-iso})
holds, one gets the corresponding identity.

For illustration and for further reference, we collect several identities for each axial symmetry
in Table~2: only the ones that are obtained by rotating the frame in coordinate planes by $90^0$.

\begin{table}[ht]
\caption{The averages that are equal to 1 for axially symmetric tensors. }
  \centering
  \begin{tabular}{|l|c@{}c@{}c@{}|}
        \hline
        \textbf{Coordinates} & & axis of  symmetry&\\
         \textbf{frame: basis}  &   &  &  \\
         \textbf{vectors} & x & y & z \\
\hline 
$({\bf e}_1,{\bf e}_2,{\bf e}_3)$ & $\langle s_1 s_2^{-2} s_3\rangle$ &
                - &   $\langle s_1^{-2} s_2\rangle$  \\
&&&\\
        $({\bf e}_1,{\bf e}_3,-{\bf e}_2)$ & $\langle s_1 m_{13}^{-1} s_3\rangle$ &
                  $\langle s_1^{-2} \sqrt{m_{13}}  \rangle$  & - \\
&&&\\
        $({\bf e}_2,{\bf e}_3,{\bf e}_1)$ & $\langle \Gamma_{22}^{-1} \sqrt{m_{23}} \rangle$ &
                $\langle \sqrt{\Gamma_{22}} m_{23}^{-1} s_3 \rangle$  & - \\
&&&\\
        $({\bf e}_2,-{\bf e}_1,{\bf e}_3)$ & - &  $\langle \sqrt{\Gamma_{22}} s_2^{-2} s_3 \rangle$ &
                               $\langle \Gamma_{22}^{-1} s_2 \rangle$  \\
&&&\\
        $({\bf e}_3,{\bf e}_1,{\bf e}_2)$ & - &  $\langle \sqrt{m_{13}} \Gamma_{33}^{-1} \rangle$ &
                     $\langle \sqrt{\Gamma_{33}} m_{13}^{-1} s_3 \rangle$  \\
&&&\\
        $({\bf e}_3,-{\bf e}_2,{\bf e}_1)$ & $\langle \Gamma_{33}^{-1} \sqrt{m_{23}} \rangle$ & - &
                $\langle \sqrt{\Gamma_{33}} m_{23}^{-1} s_3 \rangle$ \\
&&&\\
        \hline
  \end{tabular}
  %\caption{Table 1: In-text Table}
  \label{tab:2}
\end{table}

\section{ Application to velocity gradients in Navier-Stokes turbulence}

To illustrate the theory, we apply it to the results of numerical simulations of incompressible
turbulent flow. We consider two different cases: isotropic turbulence and channel flow.

The direct numerical simulation (DNS) data is provided by the Johns Hopkins Turbulence Databases
\cite{jh-iso, jh-chan}. The
%isotropic turbulence data is from the
DNS of forced isotropic turbulence was performed on a $1024^3$ periodic grid, using a
pseudo-spectral parallel code. It reaches the Taylor-scale Reynolds number that fluctuates around
$R_{\lambda}\sim 433.$ The channel flow DNS data contain a turbulent flow at a friction velocity
Reynolds number $R_{\tau}\sim1000$ within a computational domain of $L_x\times L_y\times L_z = 8\pi
h\times 2h\times 3\pi h$, where $h$ denotes the channel half-height and $x, y, z$ correspond to the
streamwise, wall-normal, and spanwise direction, respectively. The numerical grid contains $N_x
\times N_y \times N_z = 2048 \times 512 \times 1536$ grid points. Periodic boundary conditions are
applied in the longitudinal and transverse directions, and no-slip conditions at the top and bottom
walls.

\begin{figure*}[t]
    \centering
    \begin{subfigure}{0.3\textwidth}
        \includegraphics[width=\textwidth]{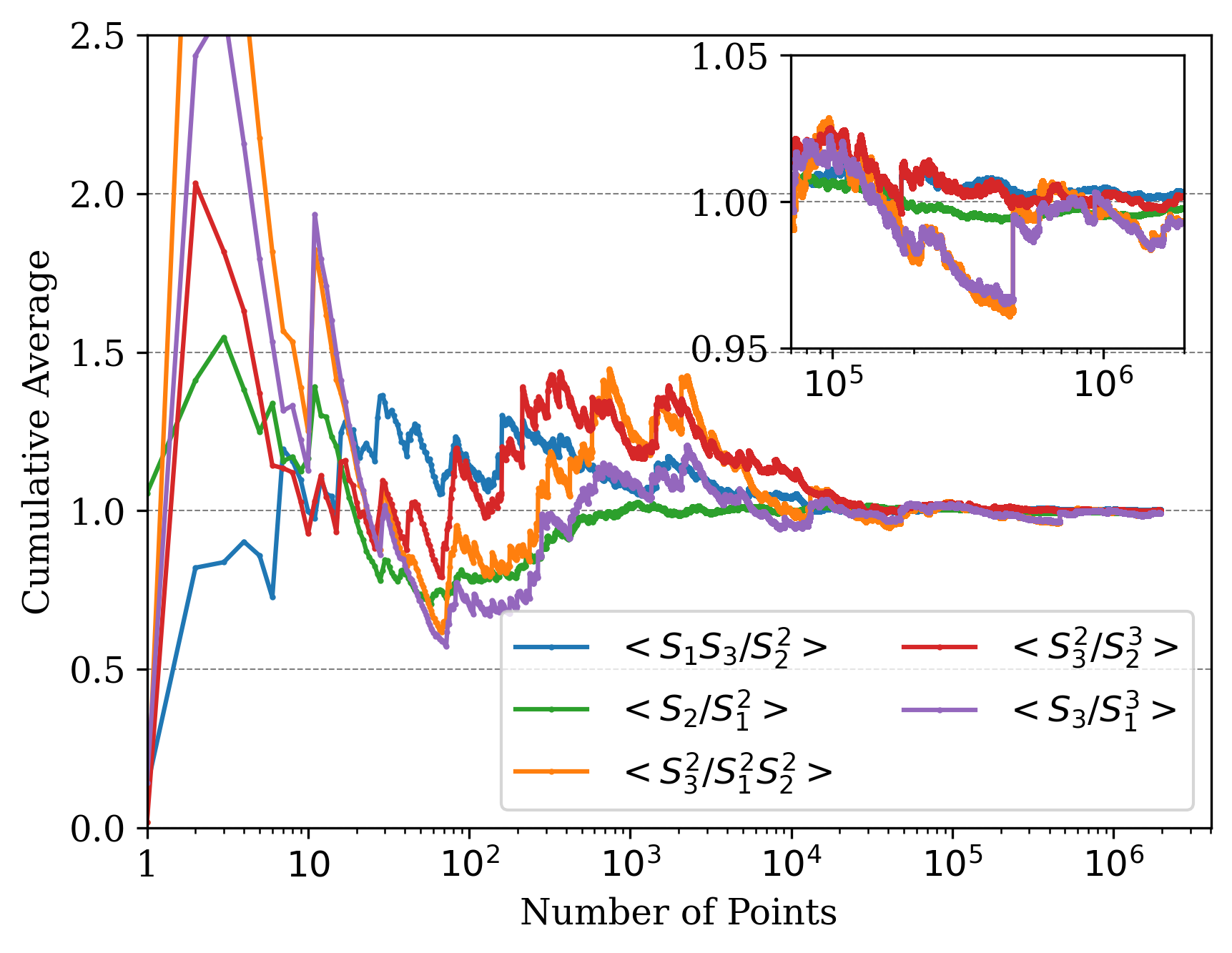}%eps}
        \caption{Isotropic flow}
        \label{fig:part1}
    \end{subfigure}
    \hfill
    \begin{subfigure}{0.3\textwidth}
        \includegraphics[width=\textwidth]{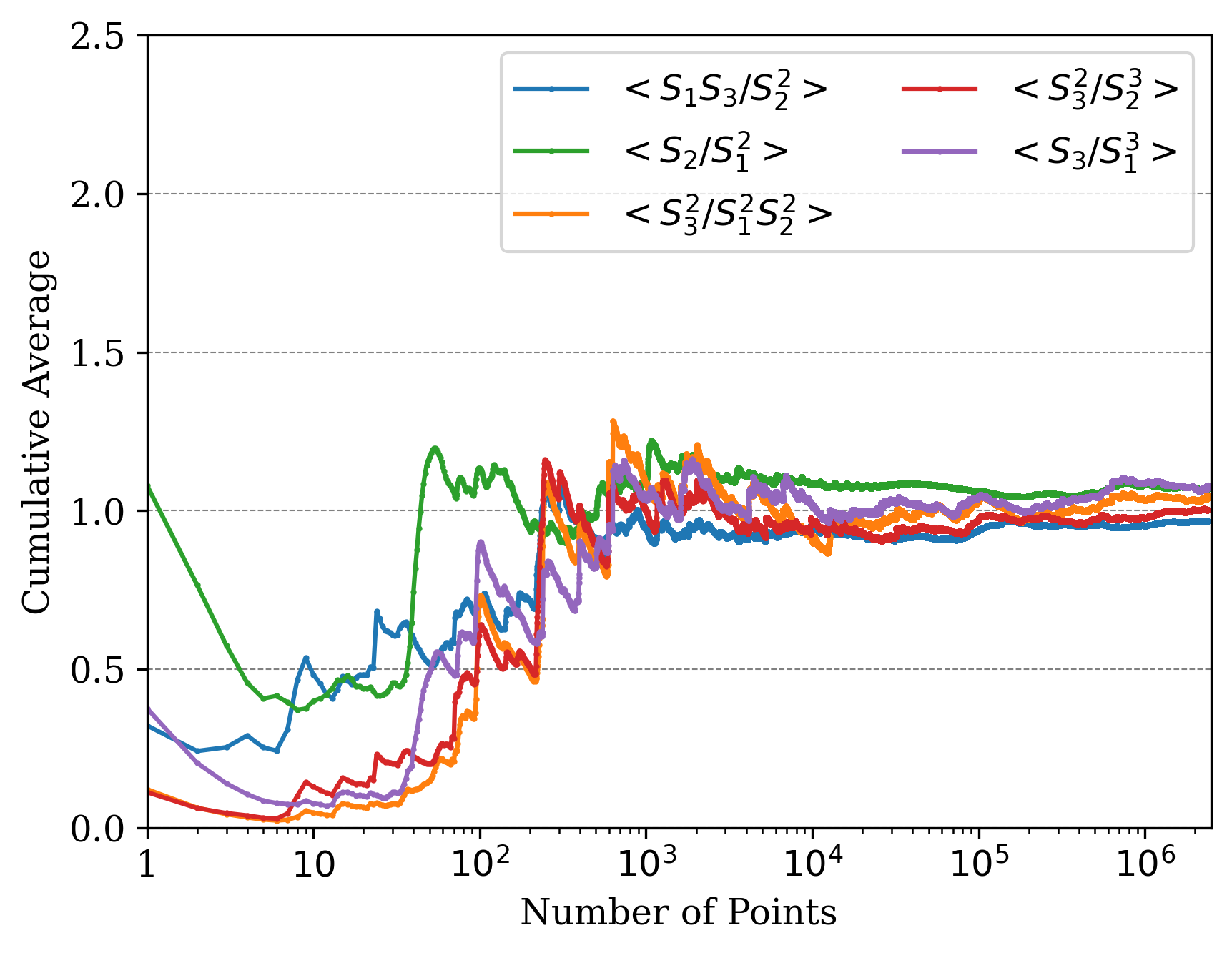}%eps}
        \caption{Center of the channel flow}
        \label{fig:part2}
    \end{subfigure}
    \hfill
    \begin{subfigure}{0.3\textwidth}
        \includegraphics[width=\textwidth]{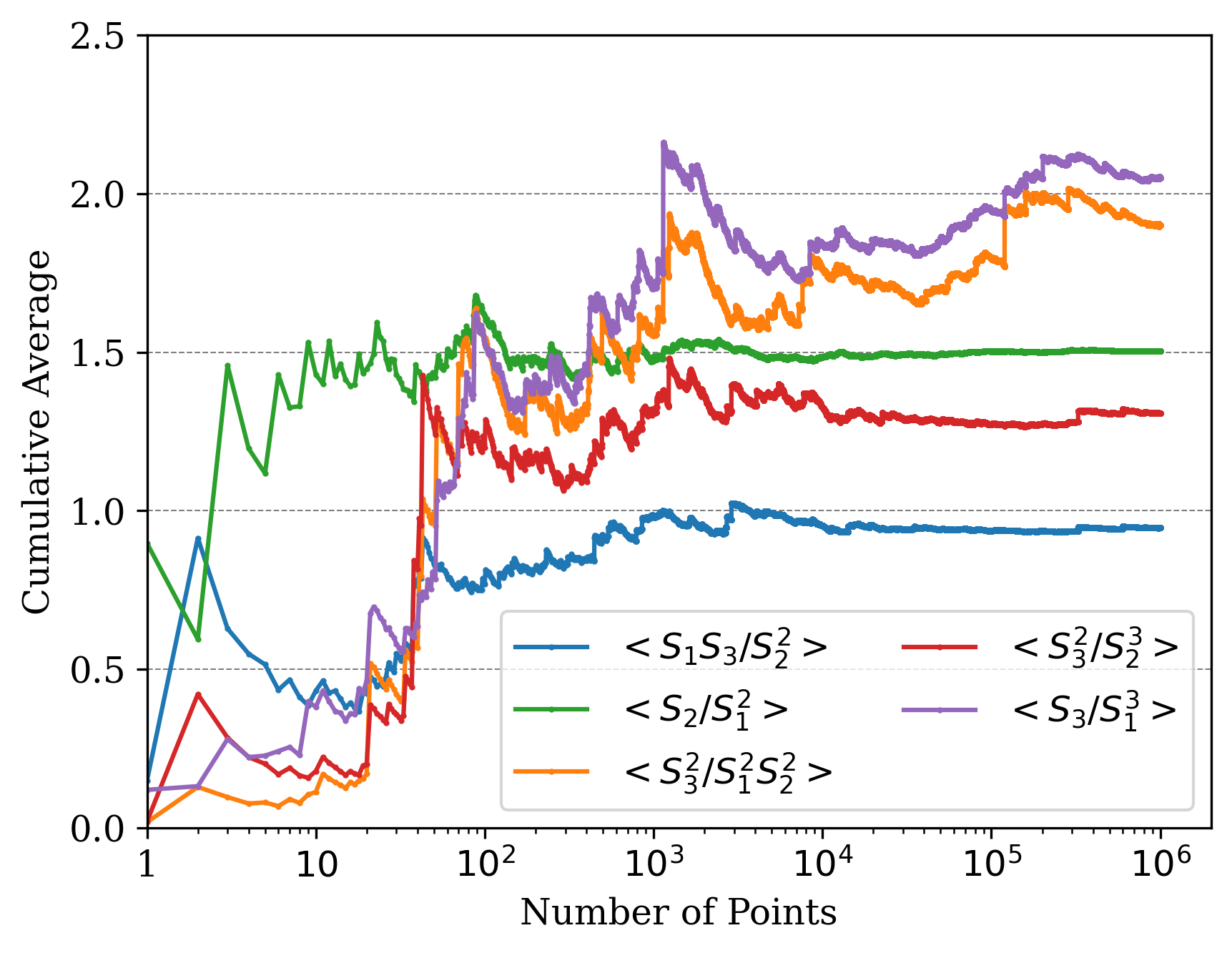}%eps}
        \caption{Vicinity of the wall, channel flow}
        \label{fig:part3}
    \end{subfigure}
    \caption{Saturation of cumulative averages for the five left-hand sides of the identities (Table
    1). The blue line corresponds to the identity that holds in the case of stochastic ($x$)-axial
    symmetry.}

    \label{fig:three_parts}
\end{figure*}

We refer the interested reader to the original publications of~\cite{jh-iso, jh-chan, jh} for a
comprehensive data description.

From the DNS we derive the velocity gradients tensor $A_{ij}=\d u_i / \d r_j$, construct the symmetric,
positive definite matrix $\Gamma=A^T A$ and calculate the left-hand sides of the identities listed
in Table~1.
%its leading principal minors.
We use the spacial average instead of the mathematical expectation,
taking advantage of %the erghodic hypothesis.
the finiteness of the correlation length.

We derive each of the averages  as a function of number of points (see Fig.1) for three different
cases: (a) isotropic turbulence ($126^3$ points at~$t=3.3$ evenly spaced in the computed domain),
 (b) centre of the channel ($2.4\times 10^6$ points in the center plane~$y=0$ at~$t=3.3$),
(c) in the channel near the wall ($10^6$ randomly distributed points in a domain near the
wall~$y\in[0.9,0.95]$ at~$t=3.3$).

One can see that, indeed, all the averages converge well to definite values; for isotropic flow the
limits of all the sequences are practically equal to unity, which confirms the identities. 
Note that  the average  $\langle s_2 / s_1^2 \rangle$, which contains small power of velocity gradient components in the denomenator,  converges very quickly, while the $\langle s_3^2 / s_1^2 s_2^2 \rangle$, which has high power of the denominator, takes  much longer time to converge.

In the
center of the channel all the averages are also close to unity; this
%indicates ... , and, thus, confirms ...
confirms that the turbulence is nearly isotropic, in accordance with the Kolmogorov
hypothesis.

%However, the difference is noticeable; only one of the identities, namely the one that
%corresponds to the axial symmetry, holds perfectly. This proves that the axial statistical symmetry
%near the center of the channel holds to very high accuracy, much better than the rest of isotropy.
As the distance from the center plane $y=0$ increases and we approach the wall, all the averages deviate
from~1. However, the axial symmetry still holds much better than the others: deviation of $\langle s_1 s_2^{-2} s_3\rangle$ {from~1}
remains at the level of less than~0.1 while the deviation of
other averages is of the order of~1.
%units (0.05 channel width from the wall). %
%\footnote{and even tens (40 at 0.025 channel width).} %

The graphs also provide an illustration of intermittency of the flow. Indeed, one can see that the
values of the averages change sharply, stepwise; these are the rare intermittent events. Moreover,
in isotropic flow %case
these rare events are essentially the ones that bring the cumulative average
back to unity: the identities, to large extent, hold due to rather rare events.

The profile of the averages as a function of the distance from the wall is presented in Fig.~2.
We add two more averages from Table II to those from Table I, to trace the possible presence of $y$-axial symmetry.
The three regions indicated in the figure correspond to the turbulent,
buffer and viscous layers. The boundaries of the zones are determined by the behavior of the mean velocity profile~\cite{jh-chan}. 

One can see that
deeply inside the turbulent layer all the averages
are close to unity, while near the buffer layer they start to
change.

The behavior of the first average in Table~1, the one that preserves in the case of  axial symmetry
around the $x$ axis, remains the most stable inside the turbulent layer.  Its deviation %of the '$x$-axial symmetric' average
from~1 remains
%about~1
quite small
even at the bound of the buffer layer,
%at the distance 0.01 channel width from the wall,
where the deviations
of the other averages are many times more.

The behavior of the averages in the viscous sublayer is close to a power-law. The two averages that reflect the symmetry aroung the $y$ axis (the axis normal to the wall) deviate from 1 much less than the others; they stop growing in the buffer layer, and remain almost constant near the wall. This indicates that, although in the viscous layer no symmetry survives,  still the symmetry with respect to the wall-normal direction violates less than the others. The value $\langle s_1^{-2} s_2^{-2} s_3^2 \rangle$ is also much smaller than the others, but it is not associated with any axial symmetry: it is merely one of the averages that must be  equal to unity in isotropic flows.  We suppose that its moderate behavior near the wall is related to the crucial role of a shear flow; below we demonstrate the arguments 
in favor of this supposition.

%To explain
To make sure that
the deviation of the averages {from~1}  
%in the viscous sublayer,
really 
%, indeed,
reflects the violation of isotropy,
one can  add small systematic anisotropy into the isotropic flow from the database~\cite{jh-iso} and watch what
happens to the averages. For example,
with this easy trick
we 'model' 
% the proximity of the wall
an effect of a systematic shear 
 by adding
constant non-random addition $\delta A_{12}=a$  
%that imitates the shear flow 
(recall that $A_{12}=\d
u_x / \d y$).  The result is presented in Fig.~3. One can see that the averages demonstrate qualitative 
 behavior similar to that   in Fig.2.
Two of them increase, and two decrease; the average $\langle s_1^{-2}  s_2^{-2} s_3^{2} \rangle$ and the two '$y$-axial' averages appear to
be the least sensitive to the symmetry violation in both Figures. The similarity of the trends may indicate that near the
wall, this 'shear' type of symmetry violation plays an important role.

\begin{figure}[t]
    \centering
    \includegraphics[width=0.95\columnwidth]{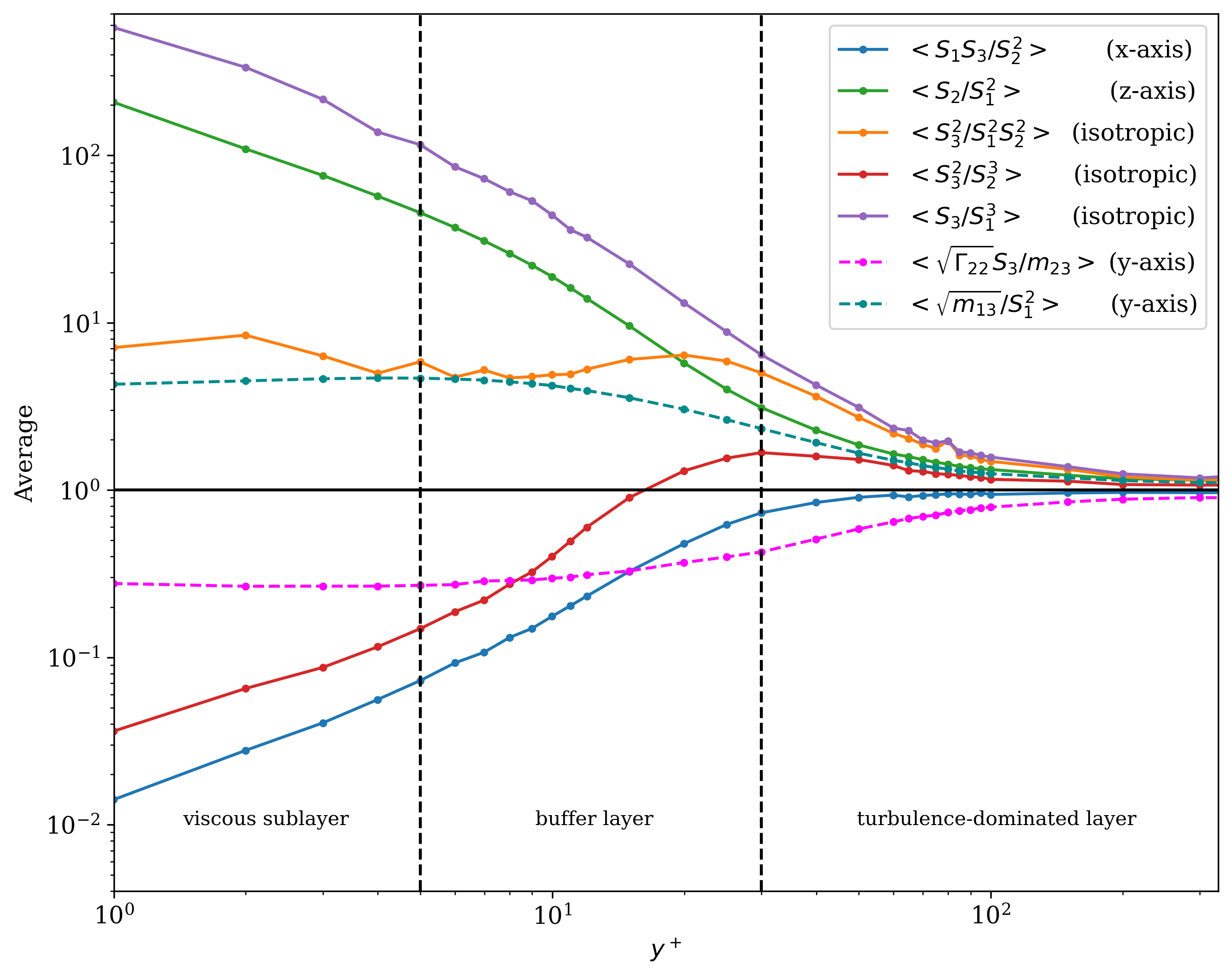}%eps}
    \caption{The averages vs the wall-normal coordinate ($y^+$). The three qualitatively different zones of a channel flow %i.e., the viscous sublayer, the intermediate buffer layer, and the turbulence-dominant zone,
    are shown by dashed lines.}
    \label{fig:part12}
\end{figure}

The sensitivity of the  averages to the distortion of isotropy in Fig.~3 is notable:
%their deviation from 1 becomes of the order of 1
they deviate from the 'isotropic' value~1 by two times, while the
%value of $a$ changes the rms
rms  value of $A_{12}$ changes quite slightly.
%only by about~10\%.
This  demonstrates the good sensitivity of the averages to the violation of statistical isotropy.

One can find more graphs describing the dependence of the averages on different components of $\bf
A$ in Supplemental Material~2: every type of addition changes the curves crucially.

We note that, apart from stochastically isotropic tensors, the identities  (\ref{E:idents}) are
also fulfilled for tensors of the form  ${\bf B} = f({\bf A}) {\bf A}$ where $\bf A$ is isotropic,
and $f({\bf A})$ is a scalar anisotropic function of $\bf A$ (e.g., $f({\bf A})$ is some numerical
function of components of $\bf A$ in some chosen coordinate frame).
\footnote{It was the Anonymous Referee who drew our attention to the fact. }
However, in real turbulence local 
anisotropy is  due to a large-scale flow, and cannot take such form.

\section{Conclusion}

\begin{figure}[t]
    \centering
    \includegraphics[width=0.95\columnwidth]{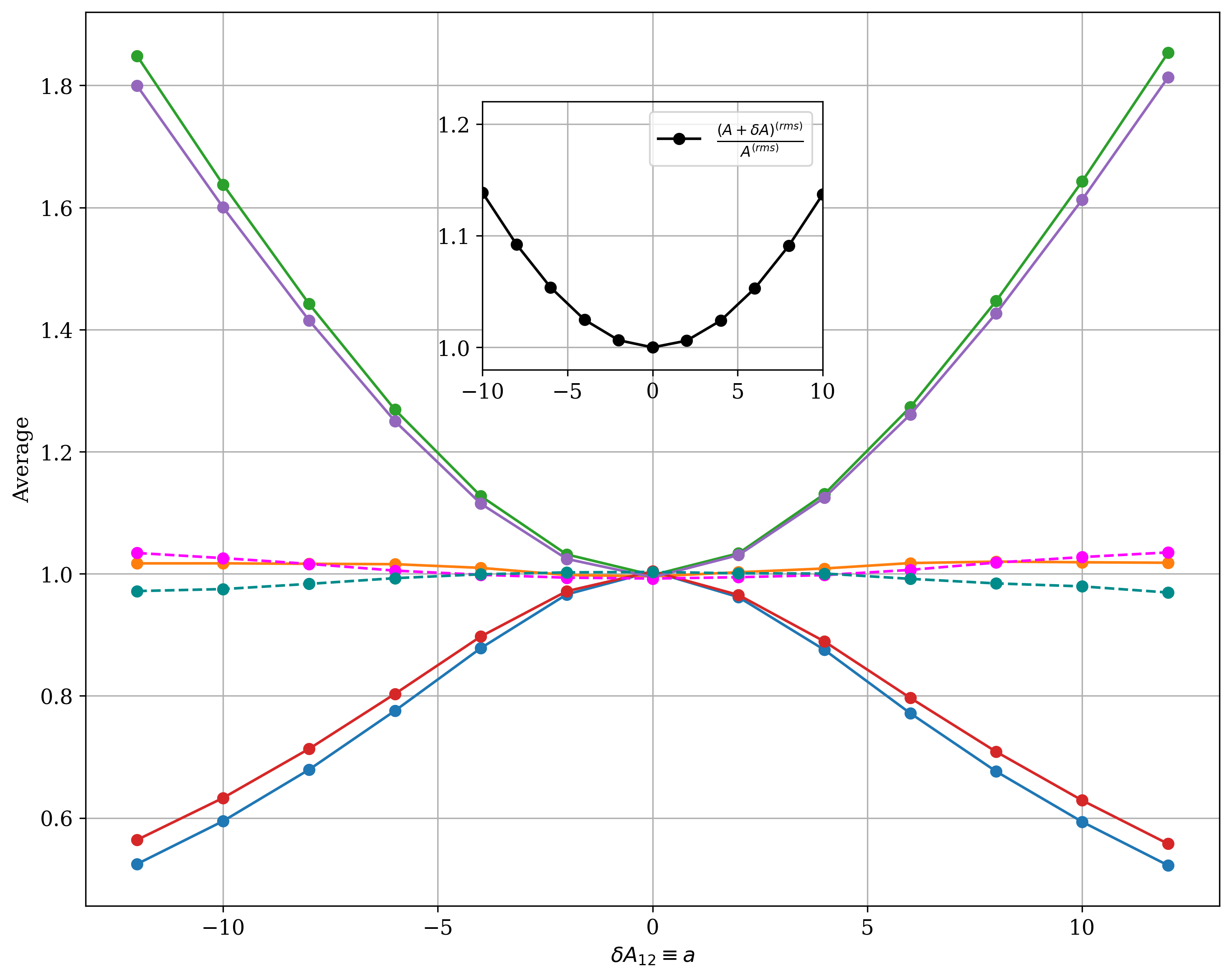}%eps}
    \caption{'Artificial shear-flow asymmetry': systematic non-random $\delta A_{12}=a$ added to the isotropic dataset for velocity gradients.
    %The value of $a$ is normalized by the inner scale of the database;
    The inset shows
    %its
    $a$
    relation to the rms value of the velocity gradient.}
    \label{fig:part22}
\end{figure}

The stochastic identities discussed in the paper are presented as the expectations of specific
dimensionless combinations of random  tensor components. If the tensor is stochastically isotropic,
all the five expectations listed in Table~1 are equal to unity; moreover, every one of them
generates a three-parametric family of identities.
% then all the averages in Table~1 are equal to unity
% as well as the corresponded to each one continuous family .
In presence of axial symmetry there are  two one-parametric families.

The discovery of these identities was made possible by the works \cite{IKSZ, SIKZ}. While
studying the evolution of coherent structures
 frozen into a random isotropic flow, we discovered a rather unexpected property of the
 evolution matrix, and  a series of new integrals of motion related to it.
 However, after further investigation it became clear that these properties are just as true for any
 stochastically isotropic tensor, not only for the evolution
  matrix.
  %\footnote{ This paper states  that the result of this symmetric property is the existence of the new non-trivial stochastic identities.}
 This paper states that the new non-trivial stochastic identities can be obtained for any
  second-rank tensor field associated with isotropic turbulent flow,
 and even for any stochastically isotropic
  second-rank tensor of any nature in a space of arbitrary dimension.

The application of the theory to the results of numerical simulations of incompressible turbulent
flow, both isotropic and in the channel, demonstrates that the averages can be used as markers of
isotropy. The analysis of the velocity
gradient tensor shows that the cumulative averages converge well to definite values,
isotropic turbulence and the center of the channel exhibiting averages close to unity, which
confirms the theory.
% As the flow approaches
Near the wall, deviations from unity become much more pronounced, with axial symmetry holding
further than other symmetries.
%showing greater stability.
 The study also highlights the sensitivity of the averages to  violation of
isotropy; this is demonstrated by  addition of small systematic anisotropy to the isotropic flow.
% , modelling the effects of wall proximity.
%
%These findings underscore the utility of the proposed method
Thus, the proposed method can be used
%in assessing the statistical properties of turbulent flows and
in investigation of turbulent flows to estimate their deviation
 from idealized isotropic conditions,  %
 the violation of the identities being treated as the measure of this deviation.

We recall that the identities discovered in the paper are valid for any kind of
tensors,  for example, for magnetic induction gradients in MHD turbulence;   and
for different types of  random fields,
including  non-stationary (in particular decaying)
turbulence.  This fact can also be validated by DNS or experiments.

 Aside from being used as the markers of isotropy, these identities can also be helpful in
 considering the theoretical problems of the turbulence theory \cite{MY75, Hill, Yakhot}.
 In this sense, we hope they can be as useful as, for instance, Ward identities have turned out to be
 in %when  creating
 renormalized quantum electrodynamics. Furthermore,%\footnote{ generalizations of the presented theory   are possible when  considering other...}
 it may be possible to generalize the presented theory for  %
 other symmetry groups, in addition to $O(d)$. Studying the symplectic group from this angle,
 for example, may be useful in stochastic Hamiltonian mechanics and similar problems. However,
 all of these ideas, while promising, are just at the beginning of their development.

\vspace{0.7cm}

We thank both anonymous Referees for valuable remarks. 
The work of A.V.K. was supported by RSF grant No. 24-72-00068.

\clearpage

\begin{widetext}

\section*{Supplemental Material}

\subsection{Geometric interpretation of the stochastic identities}

The identities derived in the paper can be described in geometric terms, in a way invariant with
respect to the choice of coordinate frame. Every one of the five continuous families of identities
then correspond to one and the same invariant geometric identity.

Consider random isotropic linear operator ${ \mathcal{A}}$ in Euclidean
space $\mathbb{R}^d$ and an arbitrary ordered set of orthonormal vectors $\{ \be_1, \be_2,\dots \be_d\}$.

The sequence of sets of vectors
$$
\be_1 \ , \ \ \{\be_1, \be_2 \} \ , \ \ \dots \{ \be_1, \be_2,\dots,\be_d \}
$$
is called a flag. It corresponds to a set of parallelepipeds of different dimensions, each one
embedded in the next one.

The action of $\mathcal{A}$ transforms the initial flag into random flag:
$$
\mathcal{A} \be_1 \ , \ \ \{\mathcal{A} \be_1, \mathcal{A} \be_2 \} \ , \ \ \dots \ \ \{ \mathcal{A}
\be_1, \mathcal{A}\be_2,...,\mathcal{A} \be_d \}
$$
We are interested in the length of the first vector $L=||\mathcal{A} \be_1||$, the square of the
parallelogram $S=||\mathcal{A} \be_1 \wedge  \mathcal{A} \be_2||$ and the rest of the hypervolumes
of the flag elements. Since $\mathcal{A}$ is a random operator, these hypervolumes are also random;
however, because of isotropy of $\mathcal{A}$, the probability density of $L$ does not depend on
the direction of $\be_1$.  Likewise, the joint probability density $\rho(L,S,\dots)$   of $L$, $S$
and other hypervolumes does not depend on the choice of the vectors $\{ \be_1, \be_2,\dots \be_d\}$.
Nor do different moments of these values depend on the initial orthonormal flag.

 If the set of vectors that generate the initial flag $\{
\be_1, \be_2,... \be_d\}$ coincides  with the standard basis,   we find out that the hypervolumes of the
resulting flag are related to the minors of the matrix representation of $\mathbf{\Gamma}=\mathcal{A}^T \mathcal{A}$ in the standard basis :
$$
L = s_1 \ , \ \  S = s_2\ , \ \ ...
$$
Thus, the five identities listed in Table 1 of the paper can be interpreted as five nontrivial
stochastic
 properties of the flag's hypervolumes.

 If the set of the flag-generating vectors does not coincide with the standard basis, the same geometric
 properties of the flag are still valid and bring new algebraic identities for any choice of the initial flag. Thus,
 rotating the initial flag (i.e.,
  the set $\{ \be_i \}$) one gets a three-parametric family of algebraic identities for each of the 
original five (or $d!-1$ identities each of them being $d(d-1)/2$-parametric in general case).
\begin{figure*}[ht]
    \centering
    \begin{subfigure}[b]{0.3\textwidth}
        \includegraphics[width=\textwidth]{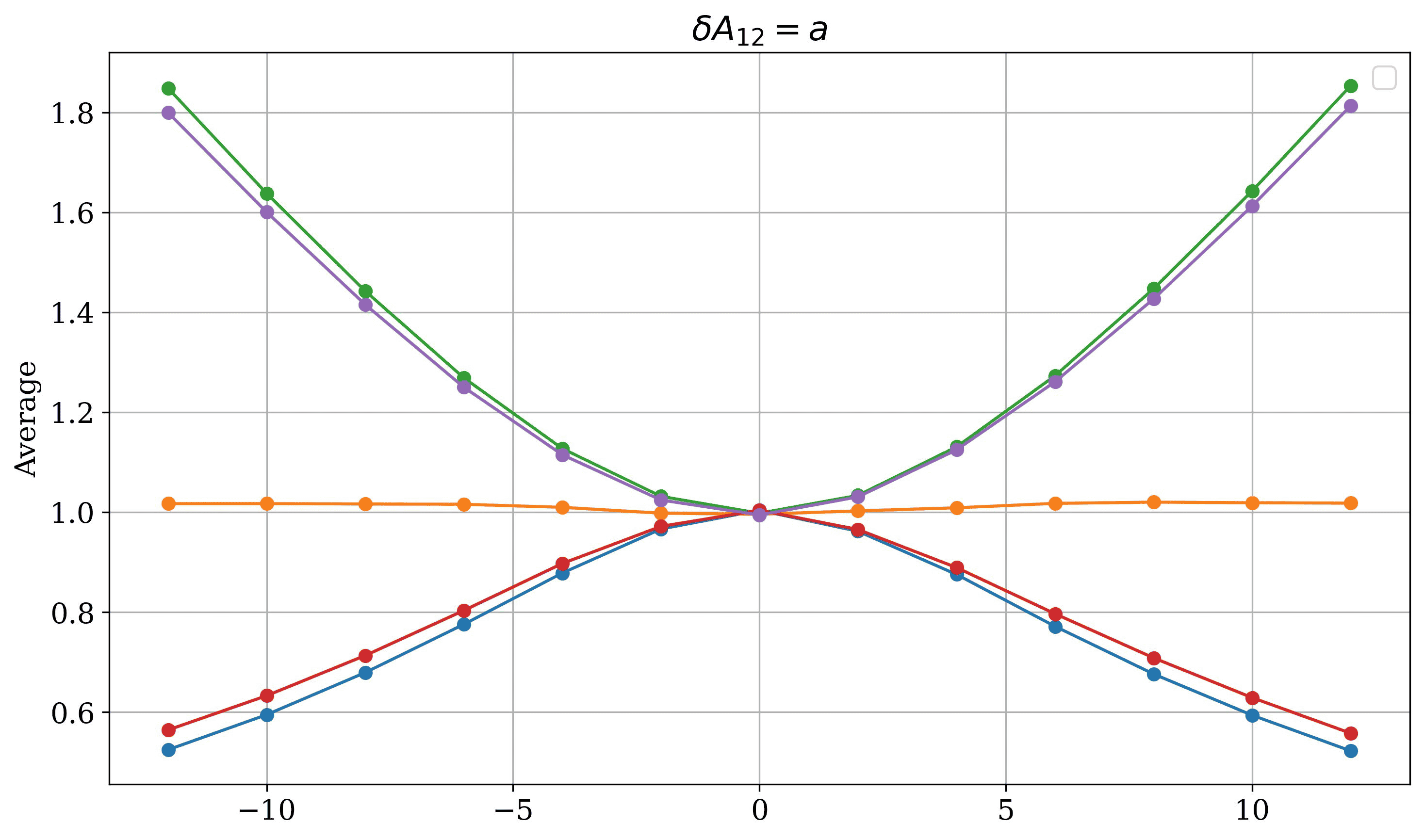}%eps}
        \label{fig:sub1}
    \end{subfigure}
    \hfill
    \begin{subfigure}[b]{0.3\textwidth}
        \includegraphics[width=\textwidth]{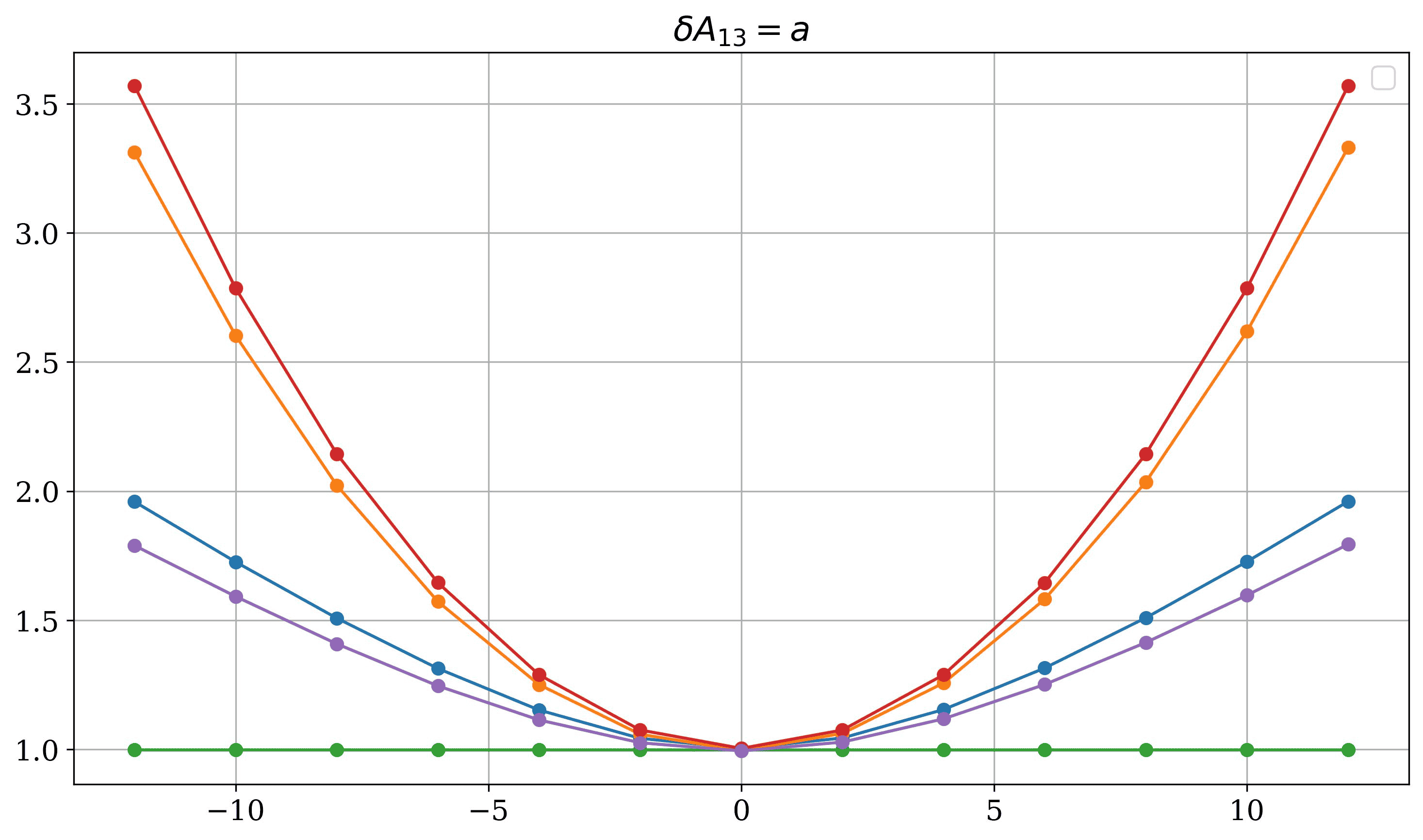}%eps}
        \label{fig:sub2}
    \end{subfigure}
    \hfill
    \begin{subfigure}[b]{0.3\textwidth}
        \includegraphics[width=\textwidth]{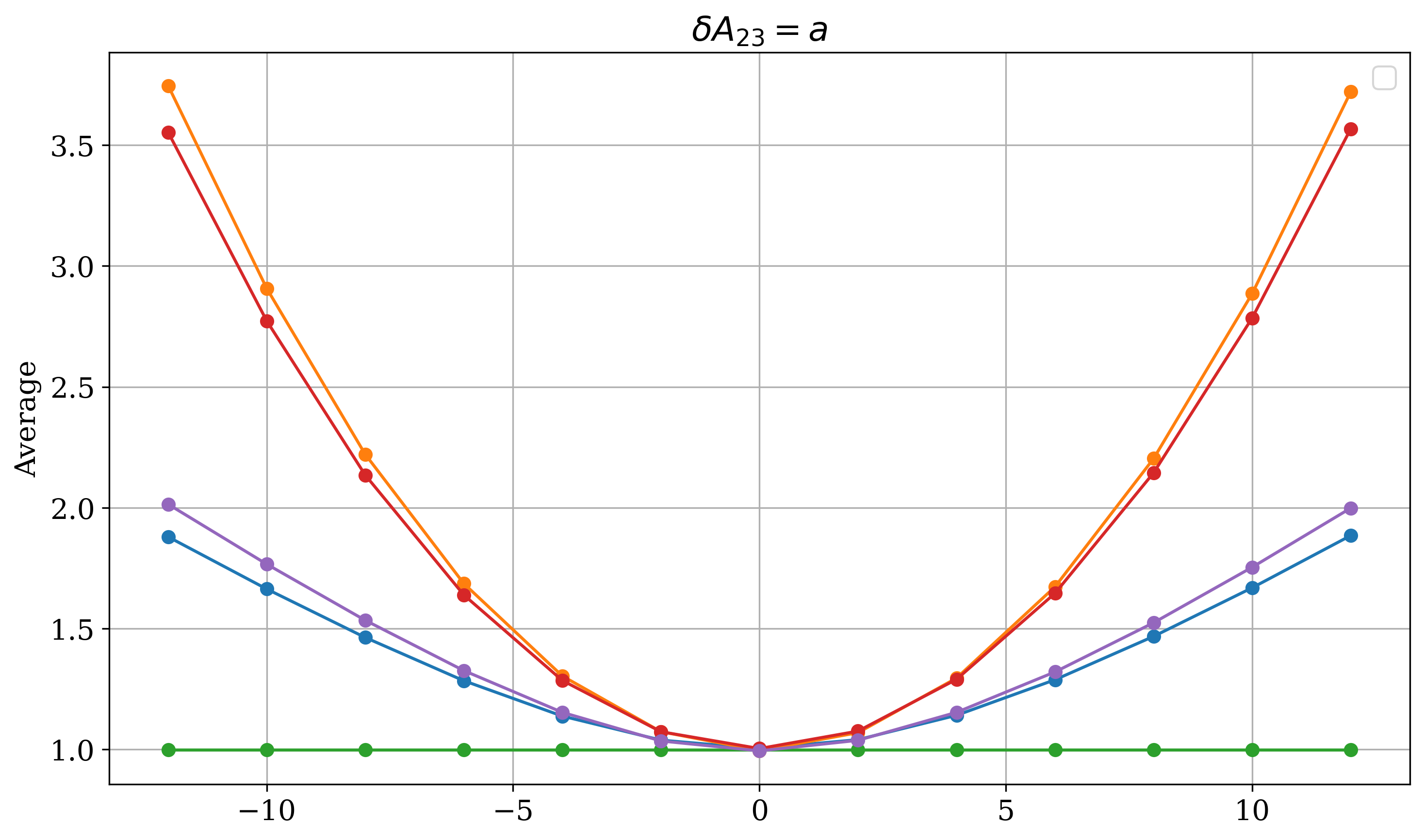}%eps}
        \label{fig:sub3}
    \end{subfigure}
    \hfill
    \begin{subfigure}[b]{0.3\textwidth}
        \includegraphics[width=\textwidth]{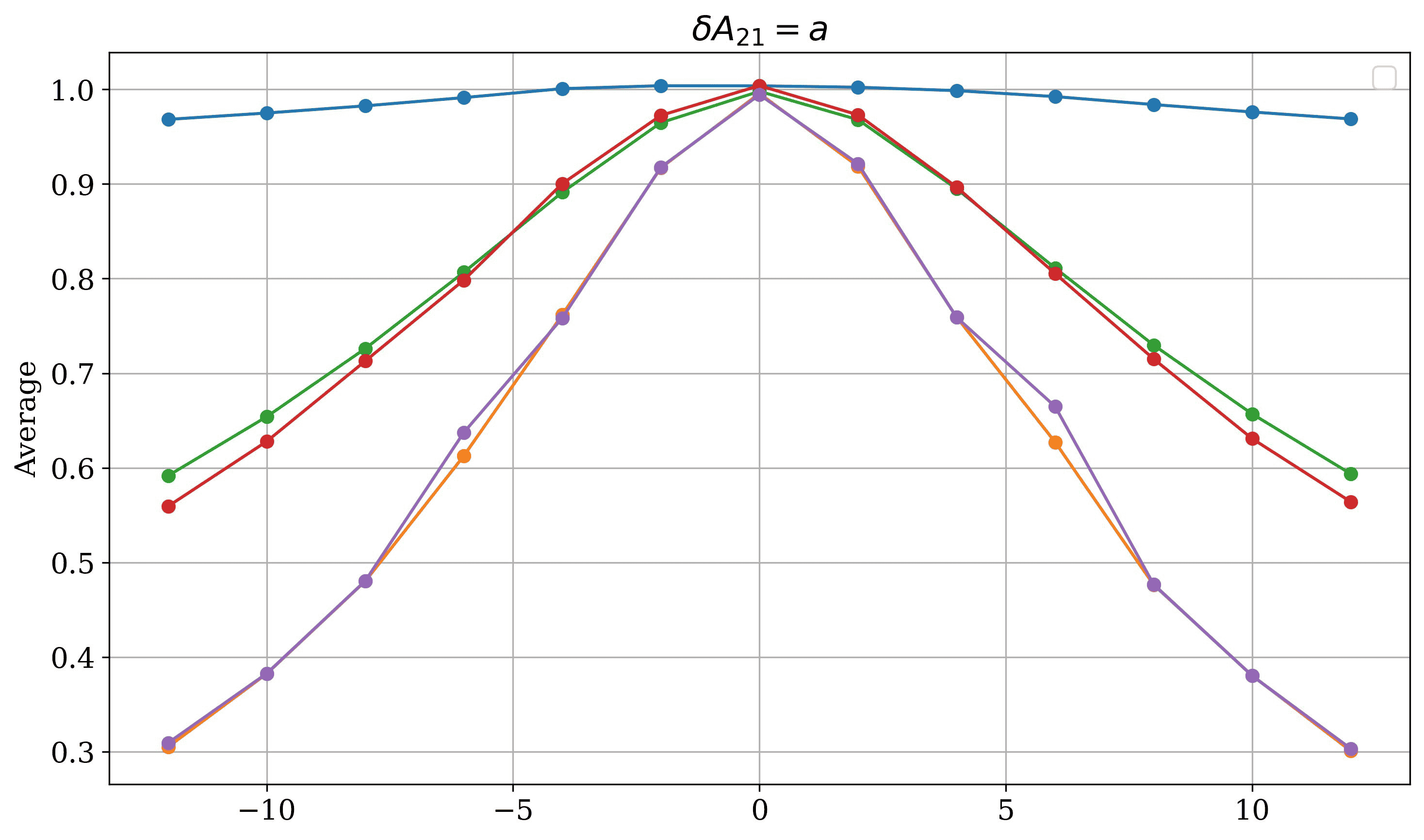}%eps}
        \label{fig:sub4}
    \end{subfigure}
    \hfill
    \begin{subfigure}[b]{0.3\textwidth}
        \includegraphics[width=\textwidth]{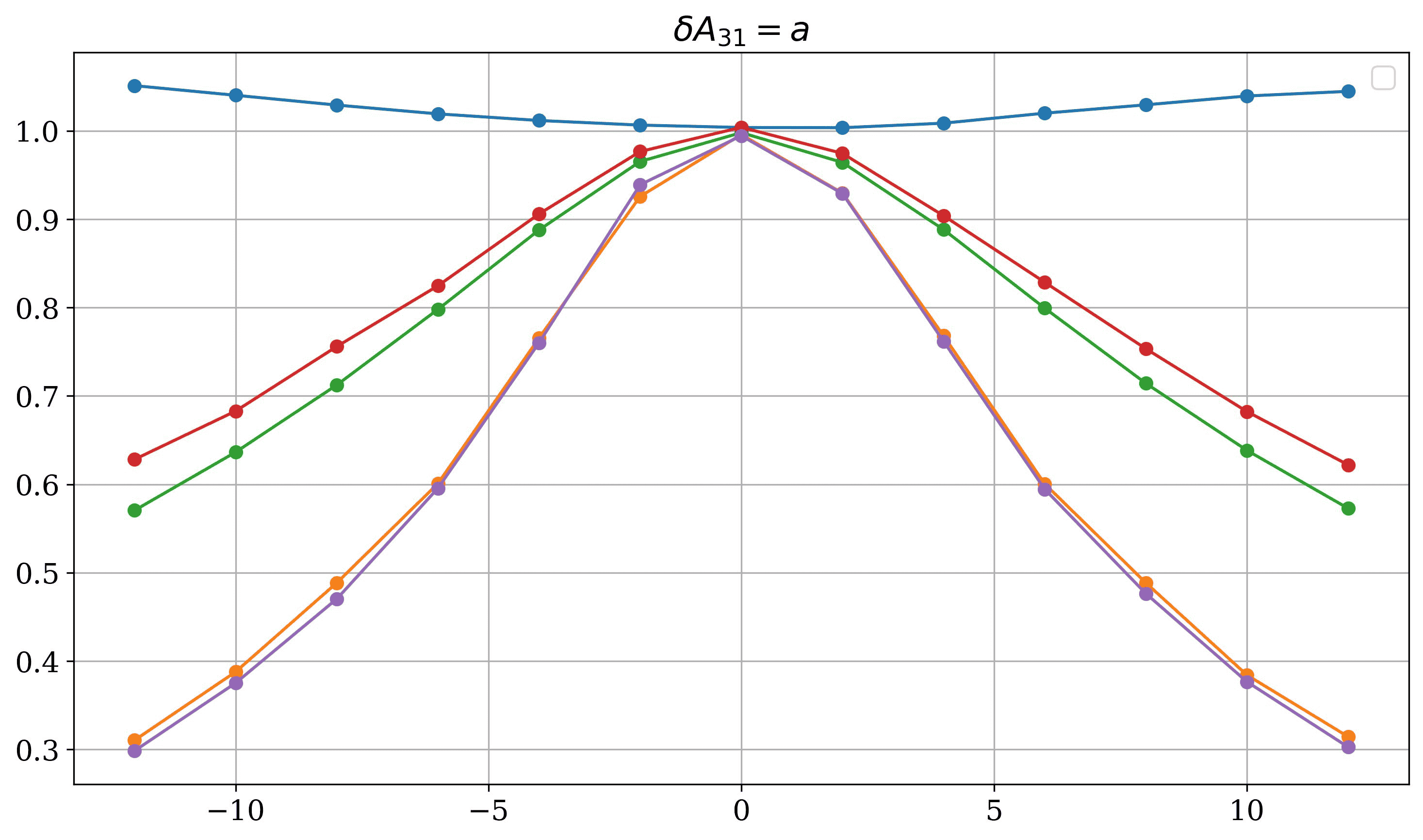}%eps}
        \label{fig:sub5}
    \end{subfigure}
    \hfill
    \begin{subfigure}[b]{0.3\textwidth}
        \includegraphics[width=\textwidth]{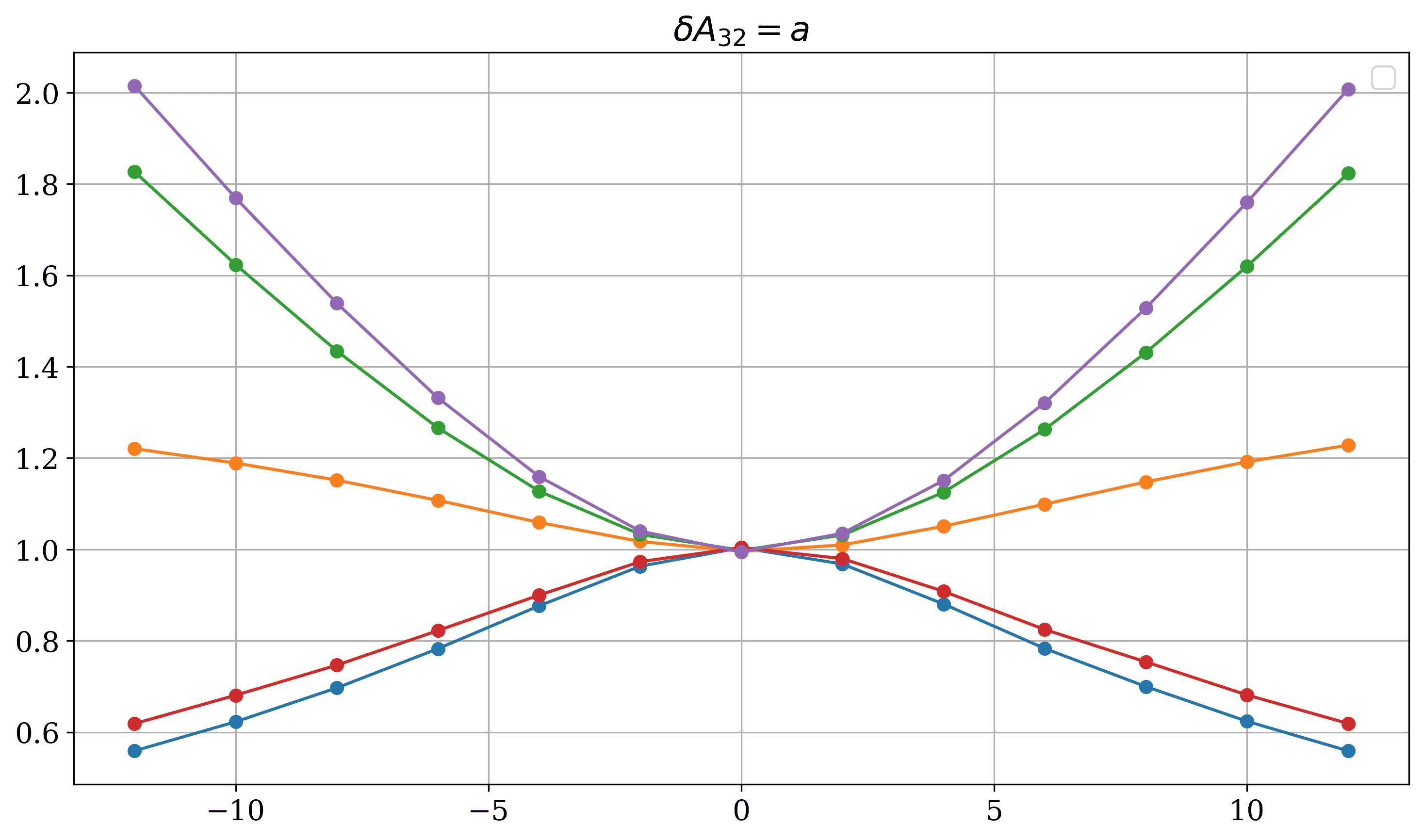}%eps}
        \label{fig:sub6}
    \end{subfigure}
    \hfill
    \begin{subfigure}[b]{0.3\textwidth}
        \includegraphics[width=\textwidth]{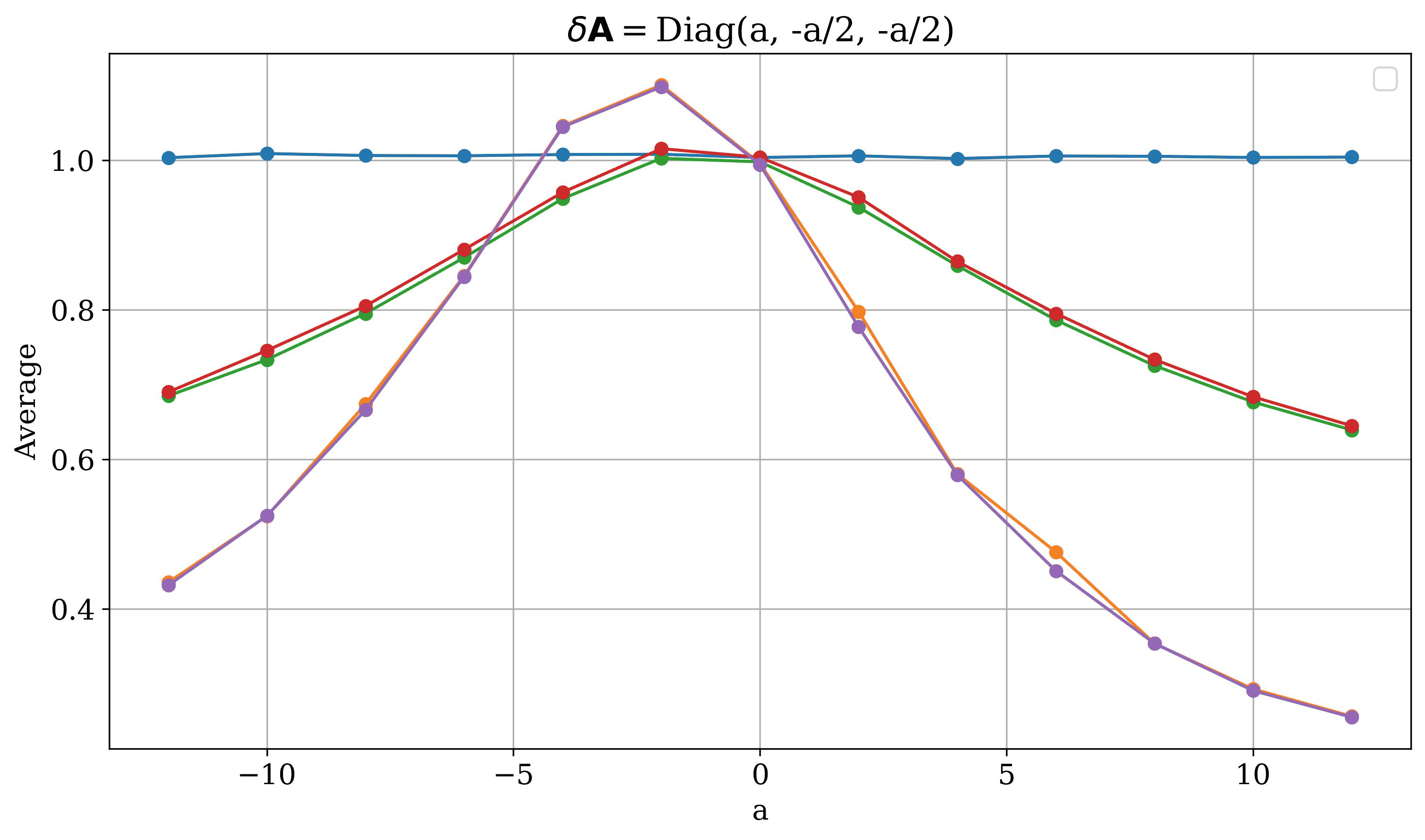}%eps}
        \label{fig:sub7}
    \end{subfigure}
    \hfill
    \begin{subfigure}[b]{0.3\textwidth}
        \includegraphics[width=\textwidth]{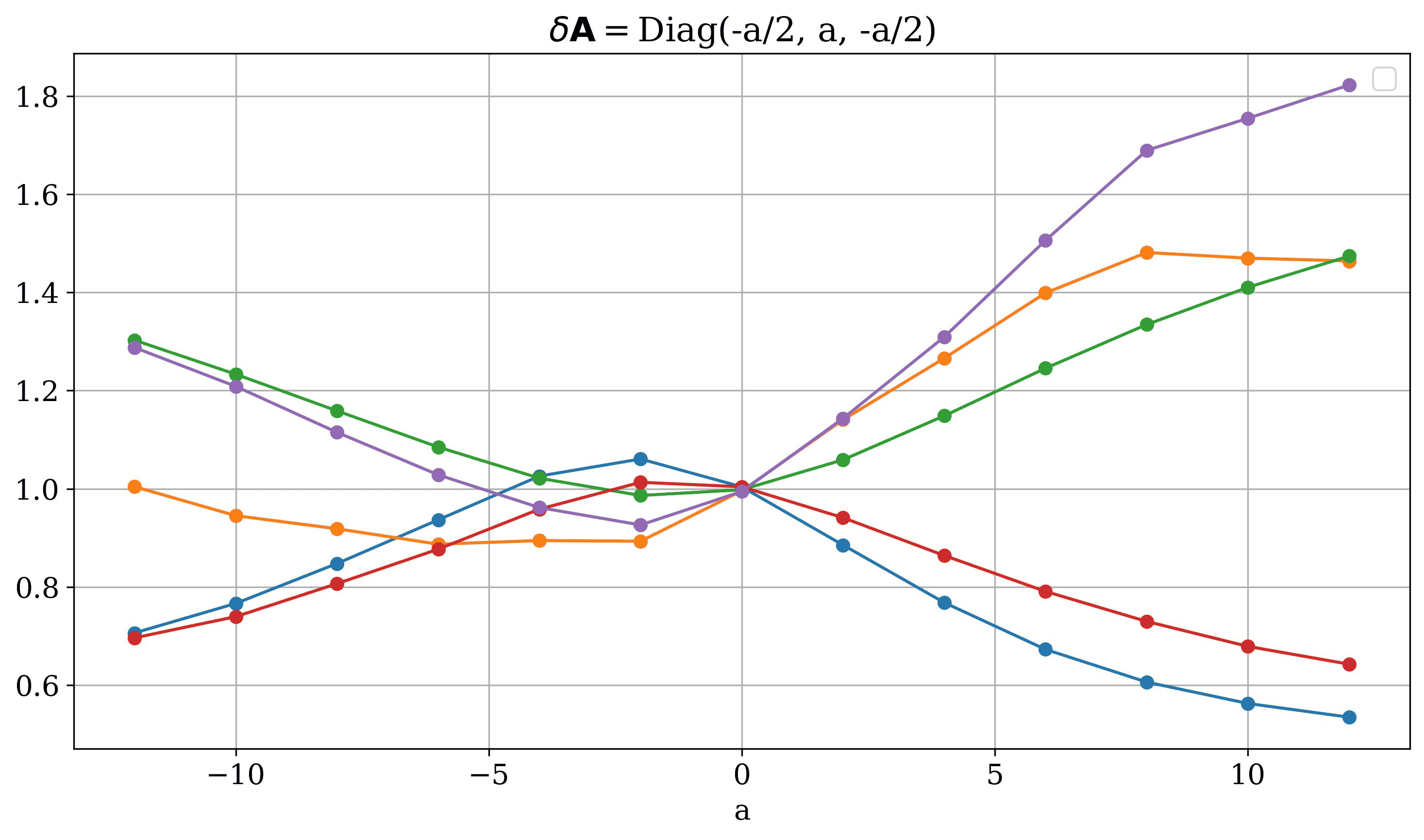}%eps}
        \label{fig:sub8}
    \end{subfigure}
    \hfill
    \begin{subfigure}[b]{0.3\textwidth}
        \includegraphics[width=\textwidth]{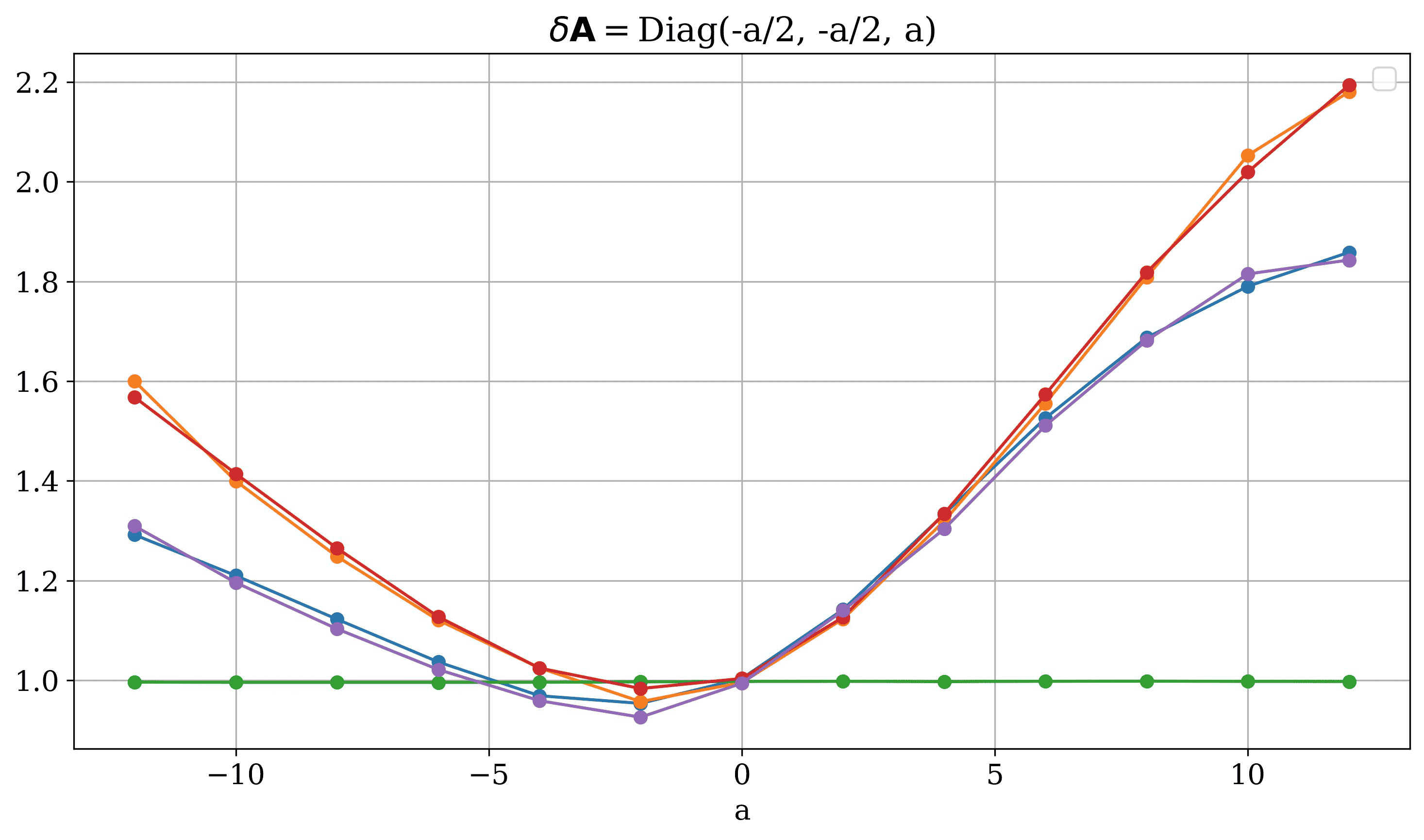}%eps}
        \label{fig:sub9}
    \end{subfigure}
    \hfill
    \begin{subfigure}[b]{0.5\textwidth}
        %\vspace{2cm} % Дополнительный вертикальный отступ для одной из картинок
        \includegraphics[width=\textwidth]{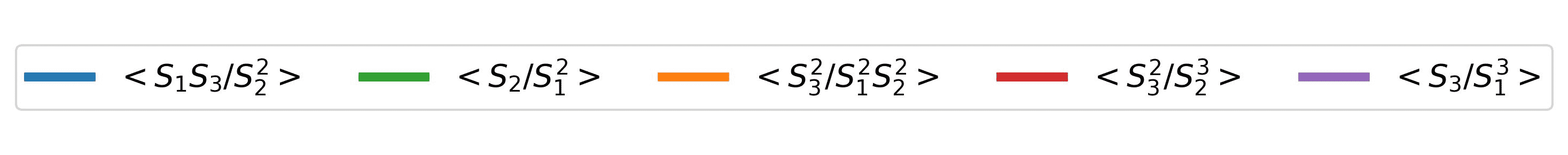}%eps}
        \label{fig:sub10}
    \end{subfigure}
    \caption{'Artificial shear-flow asymmetry': dependence of the averages on systematic non-random
    velocity gradients that distort the isotropic dataset~\cite{jh-iso-sup}.
    The type of distortion is written in the title of every graph. The legend for the averages is placed below the graphs.}
    \label{fig:grid}
\end{figure*}

\subsection{Distortion of isotropy}

To illustrate the sensitivity of the averages to different types of isotropy distortion, a series
of graphs is  presented in Figure~\ref{fig:grid}. The graphs in the first and the second row
represent % consider different
fixed shears over isotropic turbulence gradients, i.e. sequentially introduce addition of various
off-diagonal components to the velocity gradients tensor. The graphs in the third row represent
%consider
stretched incompressible cylinder over isotropic turbulent flow, i.e. introduce diagonal components
while maintaining the incompressibility condition. Note that for distortions that preserve the
% in  presence of
axial symmetry %of distortion
with respect to $x$-axis or to $z$-axis, the corresponding averages
remain unity. %even when the distortion is added.

% Дополнительный материал в одноколоночном режиме

\end{widetext}

\end{document}